\documentclass{jfm}
\usepackage{graphicx}
\usepackage{amsmath, amssymb}

\shorttitle{Transition to the ultimate regime in one-dimensional turbulence}
\shortauthor{M. Klein and H. Schmidt}

\title{The transition to the ultimate regime of thermal convection from a stochastic one-dimensional turbulence perspective}

\author{%
  Marten Klein\aff{1}
    \corresp{\email{marten.klein@b-tu.de}}
  \and Heiko Schmidt\aff{1} %
}

\affiliation{%
 \aff{1}%
  Department of Numerical Fluid and Gas Dynamics,
  Brandenburg University of Technology (BTU) Cottbus-Senftenberg,
  Siemens-Halske-Ring 14, D-03046 Cottbus, Germany
}

\begin{document}

\maketitle

\begin{abstract}ä
The Rayleigh number $Ra$ dependence of the Nusselt number $Nu$ in turbulent Rayleigh--B\'enard convection is numerically investigated for a moderate and low Prandtl number, $Pr=0.7$ and $0.021$, respectively.
Here we specifically address the case of a Boussinesq fluid in a planar configuration with smooth horizontal walls and notionally infinite aspect ratio.
Numerical simulations up to $Ra=10^{16}$ for $Pr=0.7$ and up to $Ra=8\times10^{13}$ for $Pr=0.021$ are made feasible on state-of-the-art workstations by utilising the stochastic one-dimensional turbulence (ODT) model.
The ODT model parameters were estimated once for two combinations $(Pr,Ra)$ in the classical regime and kept fixed afterwards in order to address the predictive capabilities of the model.
The ODT results presented exhibit various effective Nusselt number scalings $Nu\propto Ra^b$.
The exponent changes from $b\approx1/3$ to $b\approx1/2$ when the $Ra$ number increases beyond the critical value $Ra_*\simeq6\times10^{14}$ ($Pr=0.7$) and $Ra_*\simeq6\times10^{11}$ ($Pr=0.021$), respectively.
This is consistent with the literature.
Furthermore, our results suggest that the transition to the ultimate regime is correlated with a relative enhancement of the temperature-velocity cross-correlations in the bulk of the fluid as hypothesised by Kraichnan, R. H., \textit{Phys. Fluids}, \textbf{5}, 1374 (1962).
\end{abstract}


\begin{keywords}
Authors should not enter keywords on the manuscript, as these must be chosen by the author during the online submission process and will then be added during the typesetting process (see http://journals.cambridge.org/data/\linebreak[3]relatedlink/jfm-\linebreak[3]keywords.pdf for the full list)
\end{keywords}

\section{Introduction}
\label{sec:intro}

\begin{figure}
  \centering
  \includegraphics[width=72mm]{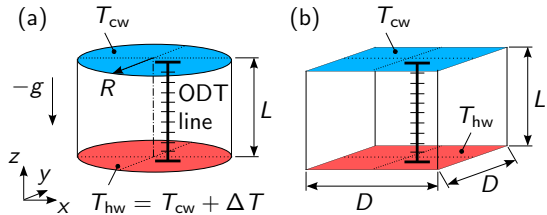}
  \caption{%
    Schematic of a (a)~cylindrical and (b)~rectangular Rayleigh-B\'enard set-up.
    The computational domain for the stochastic one-dimensional turbulence (ODT) simulations is, for both cases, a statistically representative wall-normal vertical line (ODT line).
  }
  \label{fig:setup}
\end{figure}

Rayleigh--B\'enard (RB) convection is a canonical problem for buoyancy-driven flows that are encountered in various technological and geophysical applications \citep[for an overview see][]{Chilla_Schumacher:2012}.
Typical set-ups are either cylindrical or rectangular as sketched in figure~\ref{fig:setup}. 
Fluid is confined between the heated wall ($T_{hw}$) at the bottom and the cooled wall ($T_{cw}$) at the top held at the temperature difference $\Delta T=T_{hw}-T_{cw}$.
The vertical and lateral length scales are $L$ and $D$ ($D=2R$ for a cylinder with radius $R$), respectively.
For large aspect ratios, $\Gamma=D/L\to\infty$, the flow is governed by the Rayleigh number $Ra$ and the Prandtl number $Pr$,
\refstepcounter{equation}
$$
 Ra = \frac{g\beta\,\Delta T\,L^3}{\nu\kappa},
 \qquad
 Pr = \frac{\nu}{\kappa},
 \eqno{(\theequation{\mathit{a},\mathit{b}})}
 \label{eq:RaPr}
$$
where $g$ is the constant background gravity, and $\nu$, $\kappa$ and $\beta$ are the constant kinematic viscosity, thermal diffusion and thermal expansion coefficients of the fluid, respectively.
Both $Ra$ and $Pr$ encompass several orders of magnitude in applications.
$Ra$ reaches easily up to $10^{27}$ and $Pr$ takes values in between $10^{-7}$--$10^{23}$ \citep[][]{Chilla_Schumacher:2012}.
Here we consider $Pr=0.7$ (air) and $Pr=0.021$ (mercury) for a wide range of $Ra$ numbers. 

Transitions occur when the $Ra$ number is varied \citep[][among others]{Malkus:1954b, Grossmann_Lohse:2000} and these manifest themselves in the Nusselt number,
\begin{equation}
  Nu = \frac{Q}{Q_{c}} 
     = 1 + \frac{\langle w'T' \rangle_{V,t}}{\kappa\,\Delta T/L},
  \label{eq:Nu}
\end{equation}
which is the ratio of the total, $Q$, and the purely conductive heat transfer, $Q_{c}$. 
The rightmost expression in equation~\eqref{eq:Nu} relates $Nu$ to the mean turbulent temperature flux per unit area, $\langle w'T'\rangle_{V,t}$, for the present configuration, where $\langle\cdot\rangle$ denotes a Reynolds average under statistically stationary conditions and $(\cdot)'$ the corresponding fluctuations; the subscripts $V$ and $t$ indicate domain-volume and temporal averaging, respectively,  

A transition to the ultimate regime of convection is expected for very high $Ra$ numbers when the boundary layer becomes fully turbulent \citep{Kraichnan:1962, Grossmann_Lohse:2000}.
For fixed $Pr$, the classical $Nu\propto Ra^{1/3}$ scaling \citep{Malkus:1954b} is in turn replaced by $Nu\propto Ra^{1/2}\,\left[\ln(Ra)\right]^{-3/2}$ \citep{Kraichnan:1962}.
For $Pr\simeq1$, there is evidence from laboratory experiments that the transition occurs for the critical Rayleigh number $Ra_*\simeq10^{14}$ \citep[][and references therein]{He_etal:2012, Chilla_Schumacher:2012}.
This is supported by recent two-dimensional direct numerical simulations (2-D DNSs) that have reached $Ra=10^{14}$ \citep[][]{Zhu_etal:2018}.
3-D DNSs, however, have remained in the classical regime by having reached $Ra=2\times10^{12}$ \citep[][]{Stevens_etal:2011}.
For $Pr\ll1$, the transition to the ultimate regime is expected for lower $Ra$ numbers \citep{Grossmann_Lohse:2000}.
The critical Rayleigh number is $Ra_*\simeq10^{11}$ for $Pr\simeq0.02$ \citep{Chavanne_etal:1997, Schumacher_etal:2016, Ahlers_etal:2017}.
3-D DNSs have reached $Ra=4\times10^8$, which is still in the classical regime \citep[][]{Schumacher_etal:2016}.
Unfortunately, there is no 2-D DNS data available that would support or disprove the expected transition and even the available laboratory measurements are not conclusive \citep[e.g.][]{Niemela_etal:2000}.
Complications arise because a $Nu\propto Ra^{1/2}$ scaling, or its onset, may as well be due to roughness \citep{Zhu_etal:2019} or non-Oberbeck--Boussinesq effects \citep{Urban_etal:2019}.

Therefore, the main aim of this paper is to contribute to the controversy by presenting numerical evidence for the transition to the ultimate regime of Rayleigh--B\'enard turbulence by utilising the stochastic one-dimensional turbulence (ODT) model \citep[][]{Kerstein:1999, Wunsch_Kerstein:2005}.
This lower-order modelling approach exhibits a direct turbulence cascade \citep{Kerstein:1999} and might be, in this sense, more representative for 3-D (Kolmogorov) turbulence than 2-D DNSs.
Here we specifically limit our attention to a Boussinesq fluid confined between two smooth isothermal no-slip walls.

The rest of this paper is structured as follows.
In section~\ref{sec:odt} we give an overview of the model formulation. 
In section~\ref{sec:results} we present ODT simulation results in terms of bulk profiles, Nusselt numbers, conventional temperature statistics, and temperature-velocity cross-correlations.
In section~\ref{sec:conc} we summarise and conclude our findings.

\section{Model formulation}
\label{sec:odt}

The ODT computational domain is a representative vertical line as shown in figure~\ref{fig:setup}.
The flow variables are resolved on all relevant scales along this line and evolved in time by deterministic and stochastic processes \citep{Kerstein:1999}.
For this study, we have extended the model formulation of \citet{Wunsch_Kerstein:2005} to three velocity components analogously to \citet{Kerstein_etal:2001} within the dynamically-adaptive framework of \citet{Lignell_etal:2013}.
In the following, we give a brief but complete overview of the model formulation but defer the reader to the literature for additional technical details.

\subsection{Governing equations}
\label{sec:eqs}

The governing equations are the conservation equations of mass, momentum, and energy plus an equation of state.
Here we make use of the Oberbeck--Boussinesq approximation with a linear equation of state, $\rho(T) = \rho_0\, \big[1 - \beta\,(T-T_0) \big]$, where $\rho$ is the weakly fluctuating density in addition to the other variables; the subscript $0$ denotes background values.
The density is taken as a constant except for the buoyancy forces, which yields the governing equations as \citep{Kerstein_etal:2001,Wunsch_Kerstein:2005}
\refstepcounter{equation}
$$
  \frac{\partial u_i}{\partial t} + \mathcal{E}_{i}(\alpha) = \nu \frac{\partial^2 u_i}{\partial z^2},
  \qquad
  \frac{\partial T}{\partial t} + \mathcal{E}_T = \kappa \frac{\partial^2 T}{\partial z^2},
  \eqno{(\theequation{\mathit{a},\mathit{b}})}
  \label{eq:govUT}
$$
where $(u_i)=(u,v,w)^\text{T}$ denotes the Cartesian velocity components, $t$ the time and $z$ the vertical coordinate.
Both $\mathcal{E}_{i}(\alpha)$ and $\mathcal{E}_{T}$ are stochastic terms modelling the effects of turbulent advection \citep{Kerstein:1999}.
In addition, $\mathcal{E}_{i}(\alpha)$ includes the effects of buoyancy \citep{Wunsch_Kerstein:2005} and fluctuating pressure-gradient forces \citep{Kerstein_etal:2001}, which are controlled by the model parameter $0\leq\alpha\leq1$ (see below).
Molecular diffusion is, by contrast, taken as a continuous deterministic process that is treated numerically with a finite-volume method and an explicit time-stepping scheme \citep{Lignell_etal:2013}. 

\subsection{Stochastic eddy events}
\label{sec:eddies}

$\mathcal{E}_{i}(\alpha)$ and $\mathcal{E}_{T}$ are formulated with the aid of discrete mapping (eddy) events.
Conservation and scale-locality properties are addressed by the measure-preserving triplet map \citep[][]{Kerstein:1999}.
This map compresses flow profiles over a given line interval $z_0\leq z\leq z_0+l$ (size~$l$) to one third, pastes two copies to fill the interval, and flips the central copy to ensure continuity.
Fluid is instantaneously moved from location $f(z)$ to the mapped location $z$, which yields
\refstepcounter{equation}
$$
  \mathcal{E}_{i}(\alpha):~~ u_i(z,t) \to u_i\big(f(z),t\big) + c_i(\alpha) K(z),
  \qquad
  \mathcal{E}_{T}:~~ T(z,t) \to T\big(f(z),t\big),
  \eqno{(\theequation{\mathit{a},\mathit{b}})}
  \label{eq:eddy-event}
$$
where $K(z)=z-f(z)$ is a kernel function related to the fluid displacement and $c_i(\alpha)$ are coefficients that account for momentum sources and sinks due to fluctuating pressure and buoyancy forces.
From \citet{Kerstein_etal:2001} and \citet{Wunsch_Kerstein:2005} we obtain
\begin{equation}
   c_i(\alpha) = \frac{1}{K_K}
     \left[ - u_{i,K} + \mathrm{sgn}(u_{i,K}) \sqrt{ (1-\alpha) \tilde{u}_{i,K}^2 + \frac{\alpha}{2} \tilde{u}_{j,K}^2 + \frac{\alpha}{2} \tilde{u}_{k,K}^2 } \,\right],
  \label{eq:ci}
\end{equation}
where $(ijk)$ are permutations of $(123)$ and $\tilde{u}_{i,K}^2 \equiv u_{i,K}^2 + 2 K_K\, \gamma_i\, g\beta T_K$, in which $\gamma_i\geq0$ with $\sum_{i=1}^3\gamma_i=1$ are newly introduced weights that control the conversion between the potential and the kinetic energy per component $u_{i}$.
The kernel-weighted terms follow from the literature: $K_K=\int K^2(z) \,\mathrm{d}z$ and $\phi_{K}=\int \phi\big(f(z),t\big) K(z) \,\mathrm{d}z$ for $\phi=u_i,T$.

For $T_K>0$, potential energy is released to the vertical velocity component by selecting $(\gamma_i)=(0,0,1)^\text{T}$.
When at the same time $\alpha>0$, however, some potential energy is directly transferred to the other velocity components due to implied pressure-velocity couplings.
For $T_K\leq0$, kinetic energy is extracted from all three components $u_i$.
This is constrained by $c_i$ being real-valued and addressed by selecting $\gamma_i$ as the fraction of the available kinetic energy that resides in the component $u_{i}$, that is, $\gamma_i=u_{i,K}^2\big/ \sum_{j=1}^3 u_{j,K}^2$.

Note that the original velocity vector formulation of \citet{Kerstein_etal:2001} is recovered for $g=0$ or $\beta=0$.
Likewise, the buoyancy formulation of \citet{Wunsch_Kerstein:2005} is recovered for $\alpha=0$ and $(\gamma_i)=(1,0,0)^\text{T}$ together with a single-velocity initial condition $u_0\neq0$, $v_0=w_0=0$. 
The remaining but tiny difference is the numerical evaluation of $K_K$, which, for fine grids, converges to $(4/27)l^3$ as in the references above.


A stochastic sequence of ODT eddy events aims to mimic the statistical properties of a featureless turbulent flow that, for thermal convection, is conceptually close to the assumptions made by \citet{Kraichnan:1962}.
Each eddy event is described by the random variables location $z_0$ and size $l$ for a given time $t$. 
These variables have to be sampled from the eddy-rate distribution $\lambda(l,z_0;t)=l^{-2} \tau^{-1}(l,z_0;t)$, which is general unknown. 

A thinning-and-rejection method is used in practice to avoid the construction of $\lambda$ \citep{Kerstein:1999}.
The eddy rate $\tau^{-1}$ is in turn estimated locally from the flow state, for example, by adopting a local interpretation of Prandtl's mixing length \citep{Kerstein:1999}.
Here we use an energetic formulation \citep{Kerstein_etal:2001,Wunsch_Kerstein:2005},
\begin{equation}
  \tau^{-1} \simeq C \sqrt{ \frac{1}{l^6} \Big( u_{K}^2 + v_K^2 + w_K^2 \Big) + \frac{2 K_K}{l^6}\,g\beta T_K - Z\, \frac{\nu^2}{l^4} },
  \label{eq:eddy-tau}
\end{equation}
where $C$ is the ODT eddy-rate parameter and the individual terms under the square root represent the available kinetic, potential, and a viscous penalty energy.
The latter effectively suppresses eddy events below the Kolmogorov length scale \citep{Kerstein:1999}.
Therefore, $Z$ is the ODT small-scale (viscous) suppression parameter.
Candidate eddy events are deemed unphysical and rejected when $\tau^{-1}$ is imaginary.
Otherwise they are accepted with probability $\tau^{-1}/\tau^{-1}_{s}\ll1$, where $\tau_{s}^{-1}$ is the mean sampling rate of a marked Poisson process \citep{Kerstein:1999}.
Note that no large-scale suppression method is used here so that eddy events up to the full height, $l=L$, are possible but rare.

\subsection{Model validation for turbulent Rayleigh--B\'enard convection}
\label{sec:validation}

Building on the extensive model validation by \citet{Wunsch_Kerstein:2005}, we estimated the model parameters $C$ and $Z$ for fixed $\alpha=2/3$ and $\gamma_i$ from above.
We used reference data from \citet{Stevens_etal:2011}, \citet{Li_etal:2012}, and \citet{Schumacher_etal:2016} for only two pairs $(Pr,Ra)$ as indicated in figure~\ref{fig:NuRa}.
This yielded $C=60$ for $Pr=0.7$, $C=43$ for $Pr=0.021$, and $Z=220$ for both $Pr$ \citep[][]{Klein_Schmidt:2019, Klein_etal:2018}.

\section{Results}
\label{sec:results}

In the following, the model parameters are kept fixed in order to emphasise the predictive capabilities of the ODT modelling approach.
We vary $Ra$ for $Pr=0.7$ and $Pr=0.021$, respectively, in order to address the transition to the ultimate regime.

\subsection{Bulk profiles}
\label{sec:bulk}

\begin{figure}
  \centering
  \includegraphics[width=67mm]{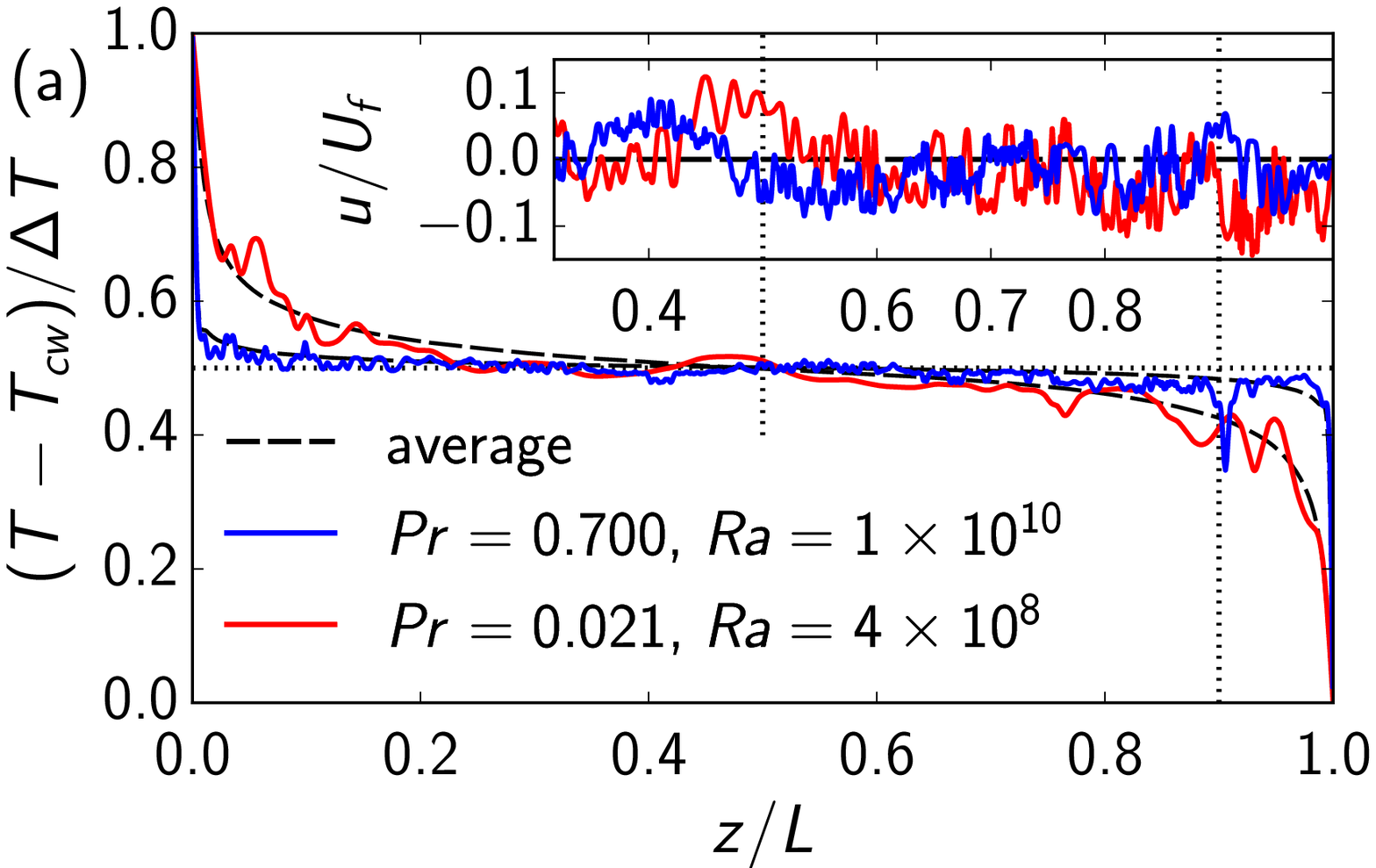} \hfill
  \includegraphics[width=67mm]{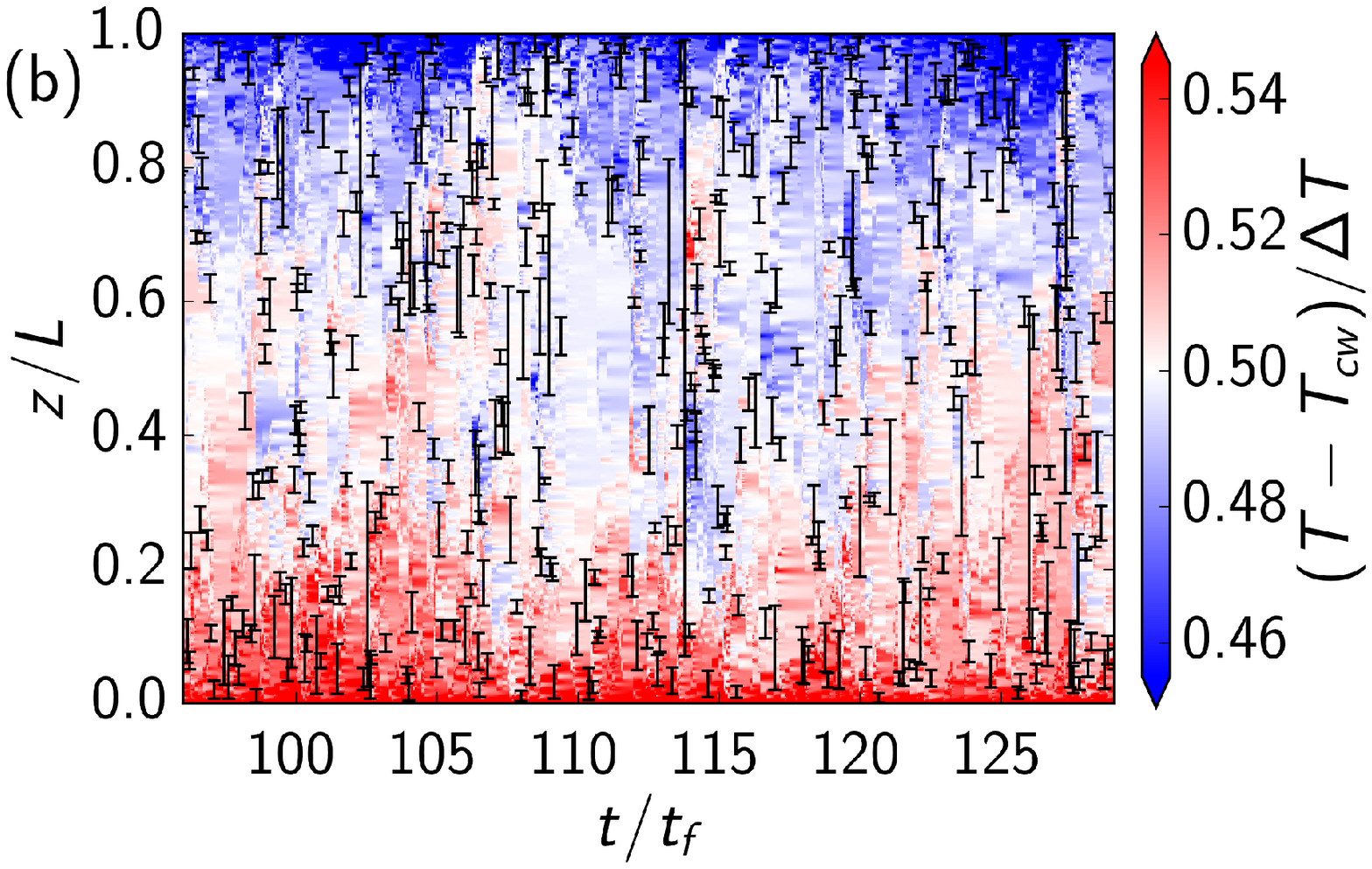}
  \caption{%
    (a)~Vertical profiles of the instantaneous, $T$, and time-averaged temperature, $\langle T\rangle$, together with the instantaneous wall-tangential velocity $u$ (inset).
    Two cases with different $Pr$ but similar $Gr=Ra/Pr=(1.6\pm0.3)\times10^{10}$ are shown.
    Dotted lines are given for orientation.
    (b)~Space-time diagram of the ODT instantaneous temperature for $Pr=0.7$, $Ra=10^{10}$.
    Black vertical lines explicitly mark every 10th instantaneous eddy event.
    $U_{f}=\sqrt{g\,\beta\,\Delta T\,L}$ is the free-fall velocity and $t_{f}=L/U_{f}$ the free-fall time.
  }
  \label{fig:bulkProf}
\end{figure}

Figure~\ref{fig:bulkProf}(a) shows profiles of the instantaneous, $T$, and the time-averaged temperature, $\langle T\rangle$, together with profiles of the instantaneous wall-tangential velocity component $u$ for the two Prandtl numbers investigated.
The cases shown exhibit approximately the same Grashof number $Gr=Ra/Pr=(1.6\pm0.3)\times10^{10}$.
The spatial scales in the instantaneous velocity fields are therefore comparable for both cases.
They are also comparable to the spatial scales in the temperature field for $Pr=0.7$ since the thermal and viscous diffusion time scales are similar.
By contrast, the spatial scales seen in the temperature field are larger for the lower $Pr=0.021$ due to faster thermal diffusion.
Furthermore, the mean temperature profiles are smooth and symmetric to the mid-height, $z/L=0.5$, which indicates well-behaved grid adaption with negligible numerical transport.
Note that the mean velocity is zero here because the mean and large-scale flow \citep[see, for example,][]{Shraiman_Siggia:1990} is not resolved by ODT.

Figure~\ref{fig:bulkProf}(b) shows a space-time diagram of an ODT temperature solution.
One can see that eddy events may displace fluid over large distances.
This can be viewed as trace of plumes in the ODT model.
The small-scale eddy events tend to follow a direct cascade, which is here likely caused by large velocity gradients on the small scales (see figure~\ref{fig:bulkProf}(a)). 

\subsection{Nusselt number}
\label{sec:nusselt}

The $Ra$ dependence of the $Nu$ number is investigated for $Pr=0.7$ and $0.021$ by stochastic ODT simulations.
The ODT model parameters are kept constant to address the predictive capabilities of the model.
$Nu$ is computed according to equation~\eqref{eq:Nu} for the 1-D computational domain by long-time averaging over several hundred thousand (low $Ra$) to several million (high $Ra$) eddy events in the statistically stationary state.
Confidence margins are obtained by computing $Nu$ in the upper and lower half of the domain and these show sufficiently converged results.
The Rayleigh numbers investigated encompass $10^{8}\leq Ra\leq10^{16}$ for $Pr=0.7$ and $10^{5}\leq Ra\leq 8\times10^{13}$ for $Pr=0.021$.
This is an extension of the high-$Ra$ preliminary results in \citet{Klein_Schmidt:2019}.

\begin{figure}
  \centering
  \includegraphics[width=67mm]{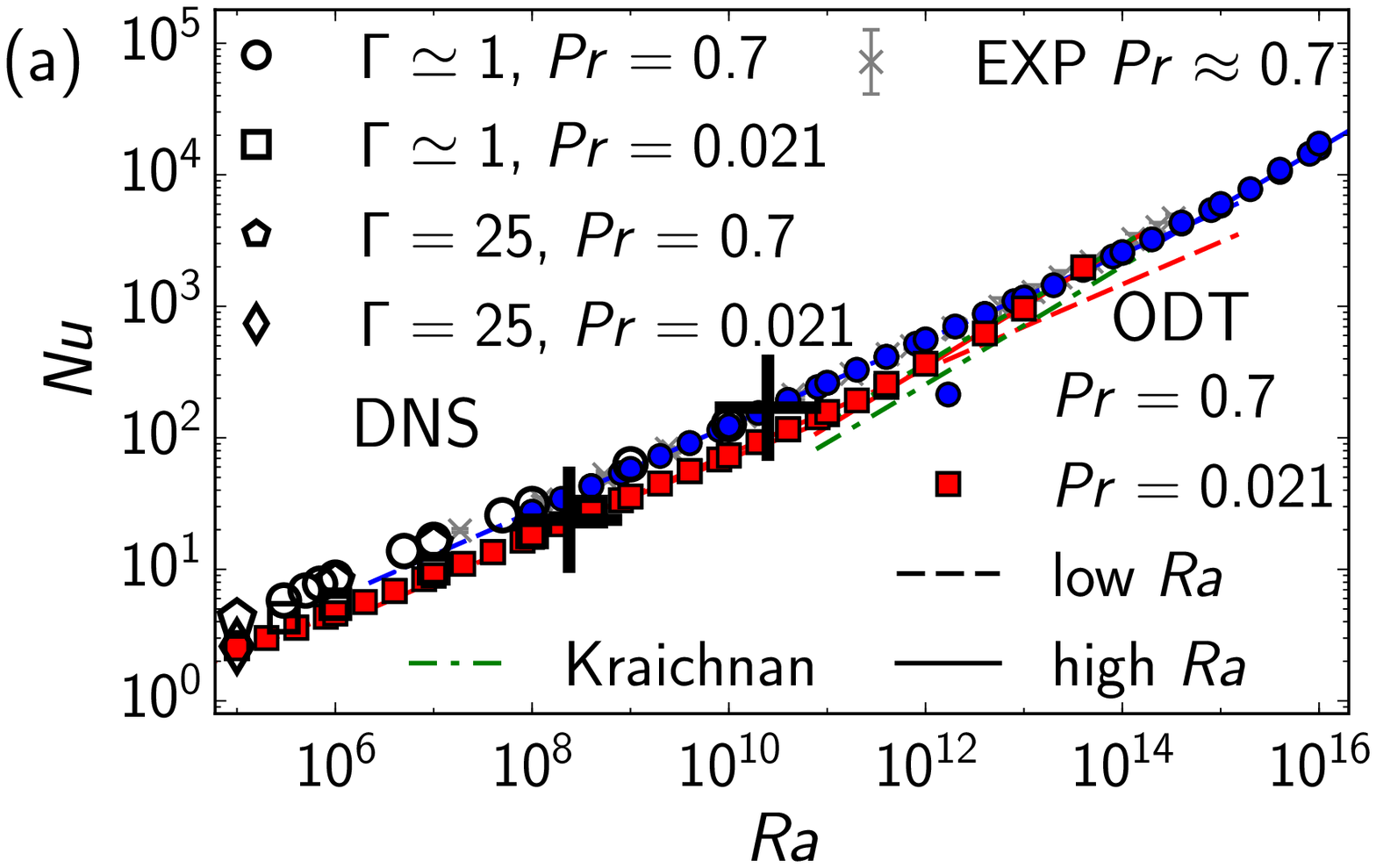} \hfill
  \includegraphics[width=67mm]{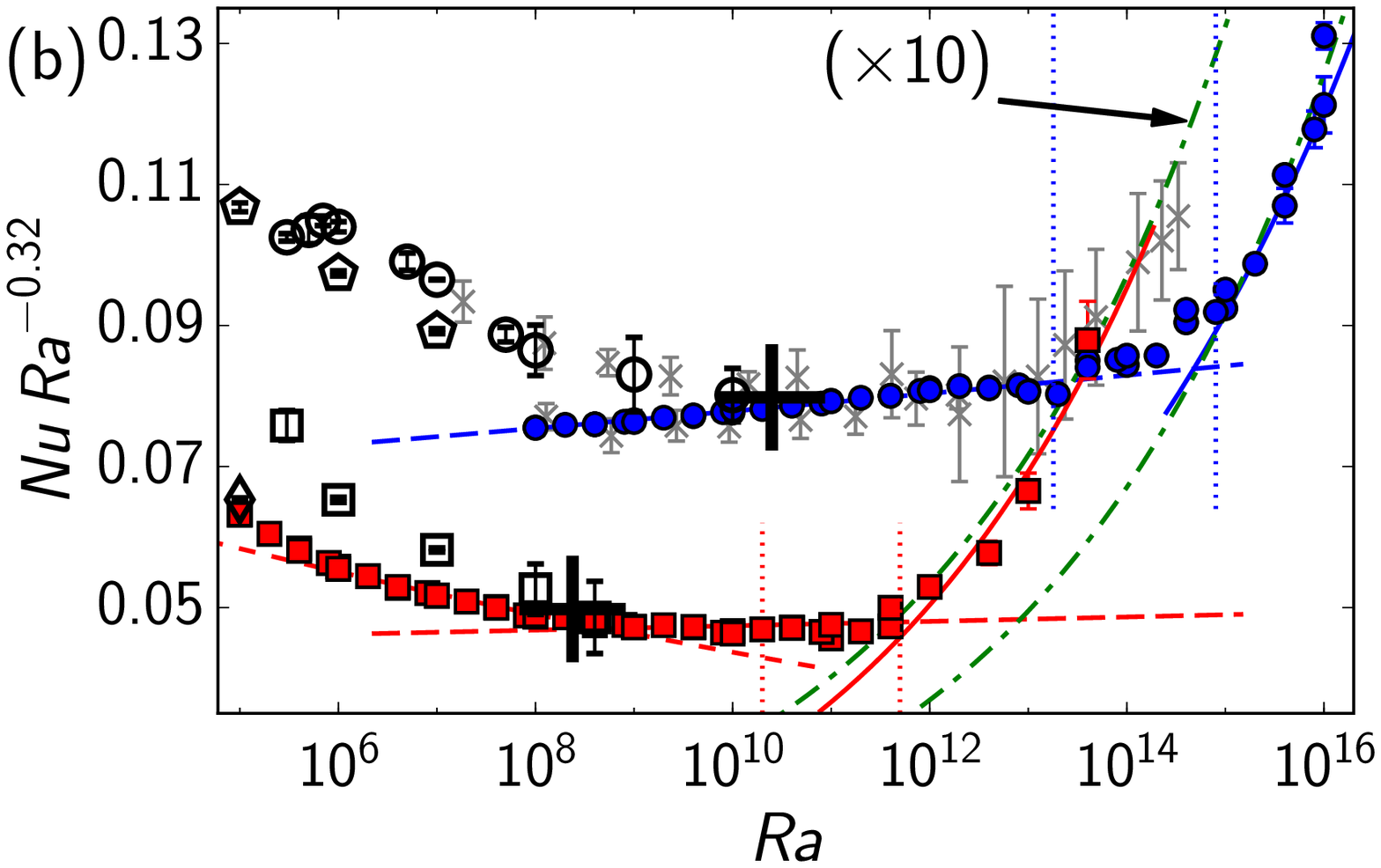}
  \caption{%
    (a)~Nusselt number $Nu$ over Rayleigh number $Ra$ for the Prandtl numbers $Pr=0.7$ and $0.021$, respectively.
    (b)~Same data but compensated with $Ra^{0.32}$.
    ODT results and the corresponding scaling laws are given.
    ODT validation cases for the $Pr$-dependent prefactor are marked by `$+$' \citep[][]{Klein_Schmidt:2019}.
    Reference DNS data are given for $1\leq\Gamma\leq3$, $Pr=0.7$ \citep{Scheel_Schumacher:2014}; $\Gamma=1$, $Pr=0.021$ \citep{Scheel_Schumacher:2016, Schumacher_etal:2016}; $\Gamma=25$, $Pr=0.7$ and $0.021$ \citep{Pandey_etal:2018}.
    Reference measurement data (EXP) encompass $0.23\leq\Gamma\leq20$, $0.5\leq Pr\leq10$ \citep[][and references therein]{Chilla_Schumacher:2012}.
    Transitional $Ra$ ranges are marked by dotted lines \citep{He_etal:2012,Schumacher_etal:2016}.
    The \citet[][]{Kraichnan:1962} scaling received an empirically magnified prefactor $(\times10)$ for $Pr=0.021$.
  }
  \label{fig:NuRa}
\end{figure}

Figure~\ref{fig:NuRa} shows $Nu(Ra)$ for $Pr=0.7$ and $0.021$, respectively.
ODT simulation results are given together with available reference data from DNS and laboratory measurements, which encompass a range of aspect ratios and $Pr$ numbers.
For very large $Ra$ numbers, the log-corrected theoretical scaling $Nu\propto Ra^{1/2}\,\left[\ln(Ra)\right]^{-3/2}$ \citep{Kraichnan:1962} serves as a reference.
The $Nu$ numbers shown in figure~\ref{fig:NuRa}(a) indicate very good agreement between the present ODT results and the available reference data across eight decades of the $Ra$ number for each Prandtl number.
All the ODT simulation runs together consumed $\approx24\,000\,\text{CPU-h}$ on local workstations equipped with Intel$^\text{\textregistered}$ Xeon$^\text{\textregistered}$ $2.40\,\text{GHz}$ CPUs.

Figure~\ref{fig:NuRa}(b) shows $Nu$ compensated with $Ra^{0.32}$ to aid the quantitative analysis.
The ODT data for fixed $Pr$ exhibit various effective scaling laws $Nu=a\, Ra^b$.
These have been obtained by a least-squares fit and are summarised in table~\ref{tbl:NuRa}.
For $Ra<10^{13}$ ($Pr=0.7$) and $Ra<5\times10^{11}$ ($Pr=0.021$), ODT exhibits the classical scaling close to $Nu\propto Ra^{1/3}$ in agreement with \citet{Malkus:1954b}.
A relative decrease of the exponent to $Nu\propto Ra^{2/7}$ \citep{Shraiman_Siggia:1990} is only observed for $Pr=0.021$, $Ra<10^9$ but not for $Pr=0.7$.
Interestingly, for $Pr=0.021$, $Ra=10^5$, ODT reproduces the reference value of \citet{Pandey_etal:2018}, which has been obtained for a large-aspect-ratio RB cell ($\Gamma=25$).  
This suggests that the ODT formulation is consistent with $\Gamma\gg1$ and is, thus, complementary to DNS and laboratory experiments with $\Gamma\lesssim1$.
We conjecture that small-aspect-ratio effects (like the unresolved large-scale circulation) together with  the energetically bounded $Ra$ range \citep[$Ra\geq 10^5\,Z\,Pr$; see][]{Wunsch_Kerstein:2005} are the main reasons for the different scaling of the ODT results observed for $Ra<10^9$.

\begin{table}
 \begin{center}
  \def~{\hphantom{0}}
  \begin{tabular}{c c c c c cl c}
    $Pr$ & $Ra_{min}$ & $Ra_{max}$ &  $a$ & ${b}$ & $a_{ref}$ && ${b}_{ref}$ \\[3pt]
    0.7~~ & $10^{8~}$ & $10^{13}$        & $0.066\pm0.001$   & $0.32\pm0.01$ & $0.15\pm0.01^\dag$ && $0.29\pm0.01^\dag$ \\
    0.7~~ & $10^{15}$ & $10^{16}$        & $(1.1\pm0.3)\times10^{-3}$ & $0.44\pm0.02$ & \;$0.78\times10^{-3\,\ddag}$ && $0.458^\ddag$ \\
    0.021 & $10^{6~}$ & $5\times10^{8~}$ & $0.078\pm0.001$   & $0.29\pm0.01$ & $0.15\pm0.04^\dag$ && $0.26\pm0.01^\dag$ \\
    0.021 & $10^{9~}$ & $5\times10^{11}$ & $0.044\pm0.005$   & $0.32\pm0.01$ & $0.45\pm0.10^*$ && $0.28\pm0.01^*$ \\
    0.021 & $10^{12}$ & $8\times10^{13}$ & $(1.5\pm0.3)\times10^{-3}$ & $0.45\pm0.02$ &  \;$1.50\times10^{-3\,\ddag}$ &$(\times10)$~ & $0.449^\ddag$ \\
  \end{tabular}
  \caption{%
    ODT scaling laws $Nu=a\,Ra^{b}$ across $Ra_{min}\leq Ra\leq Ra_{max}$ for various $Pr$.
    Reference values $a_{ref}$, $b_{ref}$ are from \citet[][$\dag$]{Scheel_Schumacher:2014, Scheel_Schumacher:2016}, \citet[][$*$]{Shraiman_Siggia:1990}, and \citet[][$\ddag$]{Kraichnan:1962}. The prefactors of the Kraichnan theory are $\tilde{a}\approx0.037$ ($Pr=0.7$) and $\tilde{a}\approx0.0054$ ($Pr=0.021$), where latter demands an empirical magnification $(\times10)$ to fit the data.
  }
  \label{tbl:NuRa}
 \end{center}
\end{table}

Moving on to high $Ra$ numbers, the transition to the ultimate regime is expected for $1.8\times10^{13}\leq Ra\leq8\times10^{14}$ \citep[$Pr=0.7$;][and references therein]{He_etal:2012,Chilla_Schumacher:2012} and $2\times10^{10}\leq Ra\leq5\times10^{11}$ \citep[$Pr=0.021$;][]{Schumacher_etal:2016, Ahlers_etal:2017}.
It is remarkable that the ODT results exhibit a transition within the expected $Ra$ ranges for fixed model parameters.
The critical $Ra_*$ numbers obtained with ODT are at the upper limits of these ranges, that is, $Ra_*\simeq6\times10^{14}$ for $Pr=0.7$ and $Ra_*\simeq6\times10^{11}$ for $Pr=0.021$, respectively.
The aspect ratio might influence $Ra_*$ indirectly by favouring some mean and large-scale motions that feed back on the boundary layer transition.
In ODT, the such motions are absent by construction.

It is remarkable also that the present ODT results are in very good agreement with the \citet{Kraichnan:1962} theory by closely following $Nu\simeq \tilde{a}(Pr)\, Ra^{1/2} \left[\ln(Ra)\right]^{-3/2}$ for the highest $Ra$ numbers investigated.
For $Pr=0.7$, Kraichnan's prefactor $\tilde{a}\simeq0.0089\,Pr^{-4}\approx0.037$ yields an almost perfect match between the asymptotic theory and ODT.
For $Pr=0.021$, however, $\tilde{a}\simeq0.037\,Pr^{1/2}\approx0.0054$ is by a factor $\approx10$ too small.
The reason for the disagreement is unclear but might be related to the fact that much higher $Ra$ numbers are targeted by the \citet{Kraichnan:1962} theory.
We emphasise here that the present ODT results are otherwise consistent with the available reference data (see below).

\subsection{Conventional temperature statistics}
\label{sec:stats}

Vertical profiles of temperature statistics have been obtained for a diagnostic grid by interpolating  and time-averaging instantaneous profiles.
We consider the non-dimensional mean temperature $\Theta$ and standard deviation of the temperature fluctuations $\sigma$,
\refstepcounter{equation}
$$
  \Theta(z) = \big(\langle T\rangle(z) - T_{b}\big) \big/ \Delta T, 
  \qquad
  \sigma(z) = \sqrt{ \left\langle T^2\right\rangle(z) - \langle T\rangle^2(z) } \big/ \Delta T,
  \eqno{(\theequation{\mathit{a},\mathit{b}})}
  \label{eq:def-Theta-sigma}
$$
where $\langle\cdot\rangle$ denotes time-averaging and $T_{b}=\left(T_{hw}+T_{cw}\right)/2$ is the bulk temperature for the present configuration (compare with figure~\ref{fig:bulkProf}).

\begin{figure}
  \centering
  \includegraphics[width=67mm]{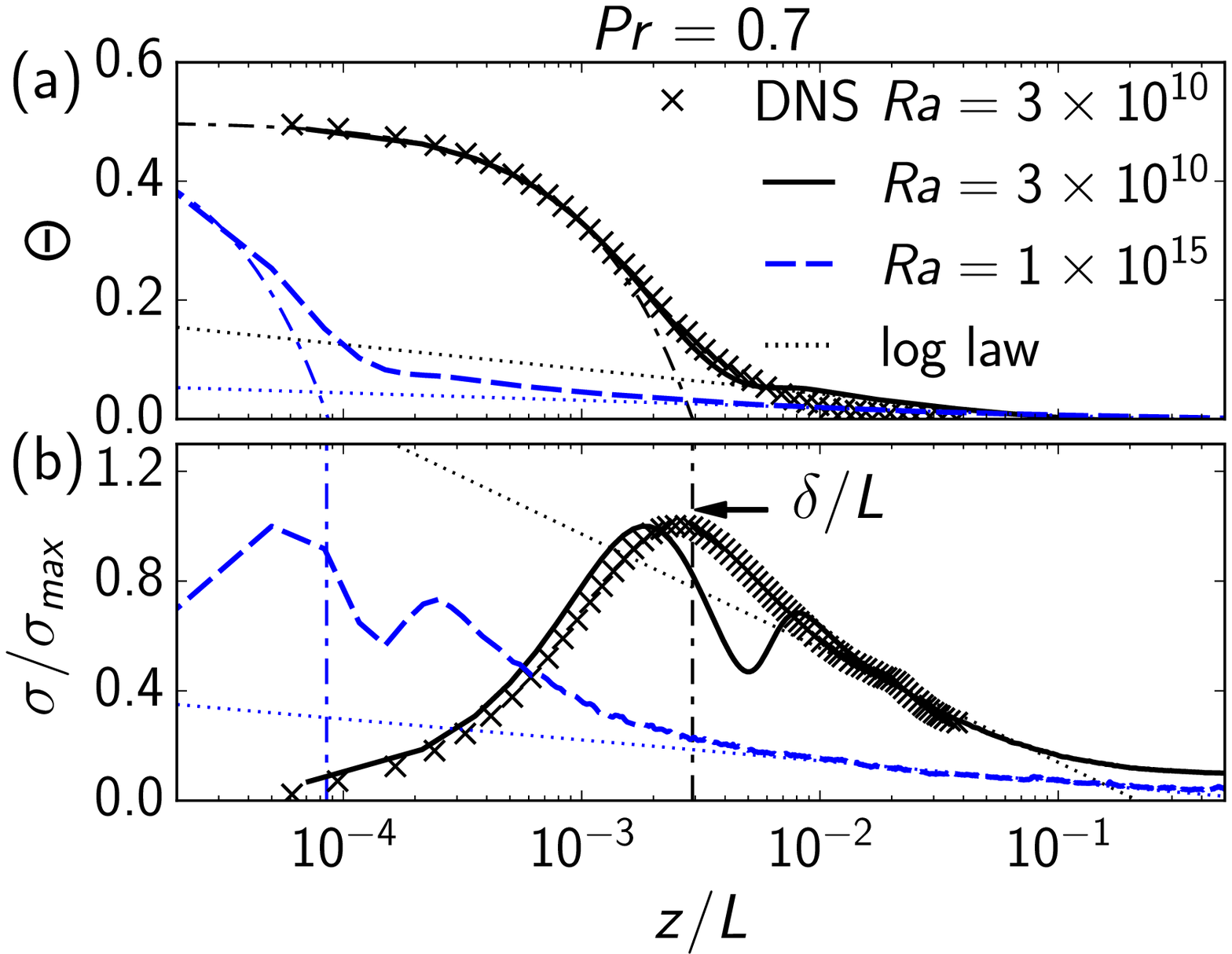} \hfill
  \includegraphics[width=67mm]{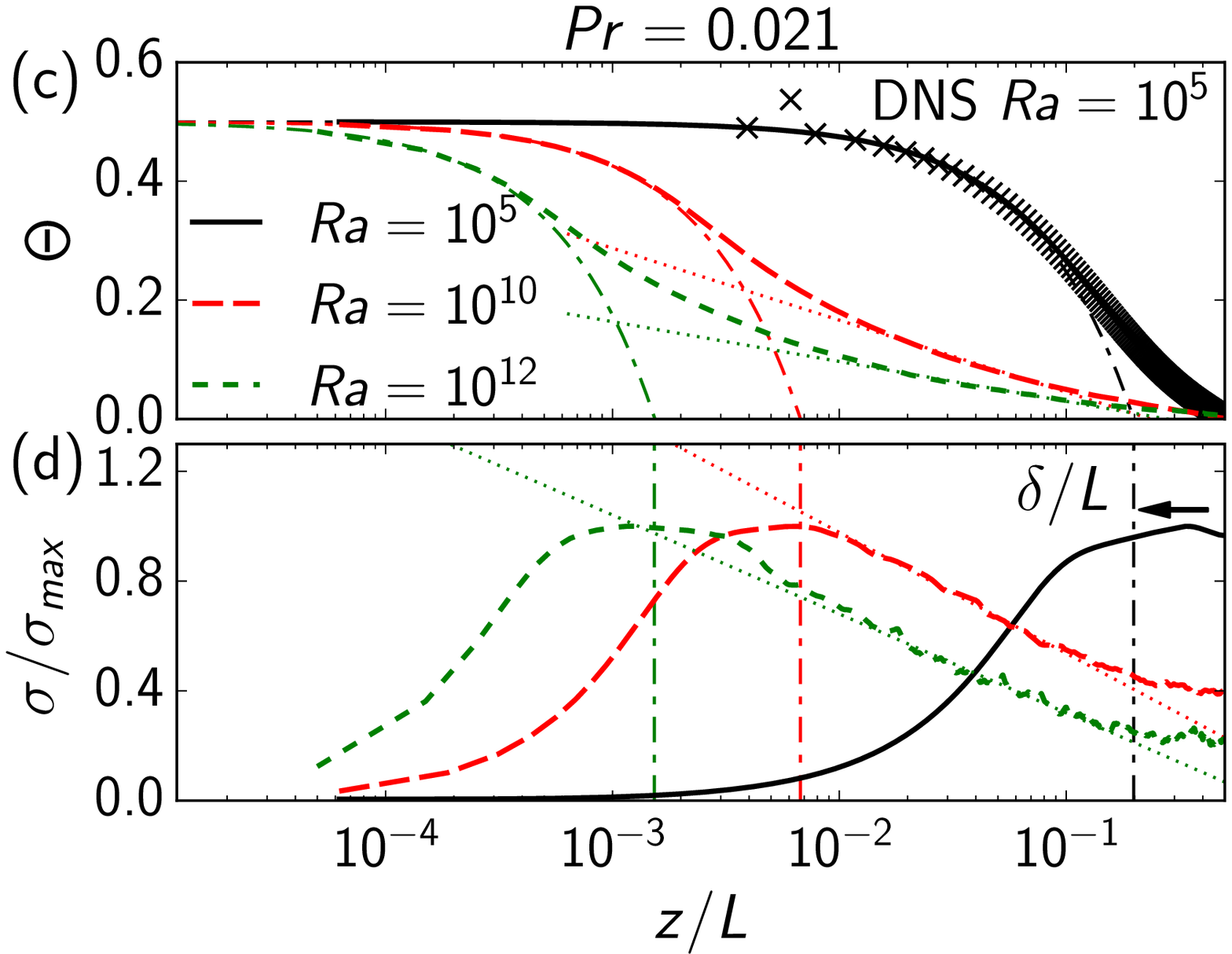}
  \caption{%
    Temperature statistics in the lower half of the domain for (a,\,b)~$Pr=0.7$ and (c,\,d)~$Pr=0.021$.
    (a,\,c)~Non-dimensional mean temperature $\Theta$.
    Dash-dotted lines give the linearly extrapolated wall gradient. 
    (b,\,d)~Standard deviation $\sigma$ of the temperature fluctuations.
    Dotted lines show equations~(\ref{eq:log-Theta-sigma}$\mathit{a,b}$) fitted to the ODT data across $10^{-2}\leq z/L\leq10^{-1}$.
    Vertical dash-dotted lines mark the thermal boundary layer thickness $\delta/L=(2\,Nu)^{-1}$.
    Reference DNS data for $Pr=0.7$ are from \citet[][]{Li_etal:2012} for the centre-line of a cylindrical set-up with $\Gamma=1$, whereas for $Pr=0.021$ they are from \citet{Pandey_etal:2018} for a rectangular set-up with $\Gamma=25$.
  }
  \label{fig:BLprof1}
\end{figure}

Figure~\ref{fig:BLprof1} shows $\Theta$ and $\sigma$ according to equations~(\ref{eq:def-Theta-sigma}\textit{a,\,b}) for $Pr=0.7$ and $0.021$, respectively.
Here $\sigma$ is normalised by its maximum value $\sigma_{max}$ in order to focus on the shapes.
Available reference DNS data are shown for comparison.
The flow statistics are symmetric to the mid-height so that we present data only for the lower half of the domain ($z/L\leq0.5$).
At the wall, the heat transport is entirely carried by molecular diffusion so that $Nu$ is related to the wall-temperature gradient, $Nu=-(\text{d}\langle T\rangle/\text{d}z)_{hw}\big/(\Delta T/L)$.
This yields the thermal boundary layer thickness as $\delta/L=(2\,Nu)^{-1}$, which is given in figure~\ref{fig:BLprof1} and indicates that the resolution is high enough to resolve all relevant features.

In general, $\Theta$ and $\sigma$ are well-captured by ODT in the vicinity of the wall and further towards the bulk for both $Pr$ numbers investigated.
For $Pr=0.7$ in figures~\ref{fig:BLprof1}(a,\,b), one can discern a spurious, undulating structure in the ODT results for finite distances from the wall, $2\times10^{-3}\lesssim z/L\lesssim10^{-2}$.
This feature is a modelling artefact, which is related to the triplet map \citep{Lignell_etal:2013}.
Interestingly, this feature has disappeared for $Pr=0.021$ in figures~\ref{fig:BLprof1}(c,\,d).
We attribute this effect to the the larger thermal diffusivity in the case of a lower $Pr$ number.
The turbulence is also more vigorous and, considering the velocity field, exhibits a broader range of scales.
So, not only molecular but also turbulent processes contribute to diffuse the imprint of the near-wall self-similar mapping.
This is, in fact, consistent with the effects observed for a smaller ODT viscous suppression parameter $Z$ \citep[see][]{Klein_etal:2018}. 

The ODT simulated profiles $\Theta(z)$ and $\sigma(z)$ exhibit a logarithmic region in the classical and ultimate regime when the $Ra$ number is large enough.
The profiles take the form
\refstepcounter{equation}
$$
  \Theta(z) = A \,\ln\left(z/L\right) + B,
  \qquad
  \sigma(z) = C \,\ln\left(z/L\right) + D,
  \eqno{(\theequation{\mathit{a},\mathit{b}})}
  \label{eq:log-Theta-sigma}
$$
where the coefficients $A,B,C,D$ are obtained by a least-squares fit across $10^{-2}\leq z/L\leq10^{-1}$.
These fits are shown as dotted lines in figure~\ref{fig:BLprof1}, where only $Ra=10^5$ for $Pr=0.021$ has been excluded as it does not exhibit a logarithmic region.
The coefficients $A$ and $C$ approach zero from below with increasing $Ra$ number.
All of these aspects are in very good qualitative agreement with \citet{Ahlers_etal:2012}.

\subsection{Temperature-velocity cross-correlations}
\label{sec:flux}

The temperature-velocity cross-correlations $\langle w'T'\rangle$ are computed in ODT on the basis of the map-induced changes \citep{Kerstein:1999}.
Cross-correlations are therefore rather well captured even when the auto-correlation of the fluctuations themselves is underestimated \citep[][]{Kerstein:1999, Kerstein_etal:2001, Klein_etal:2019}.
We make use of this property of the model to gain further insight into the boundary layer for different convection regimes.

\begin{figure}
  \centering
  \includegraphics[width=67mm]{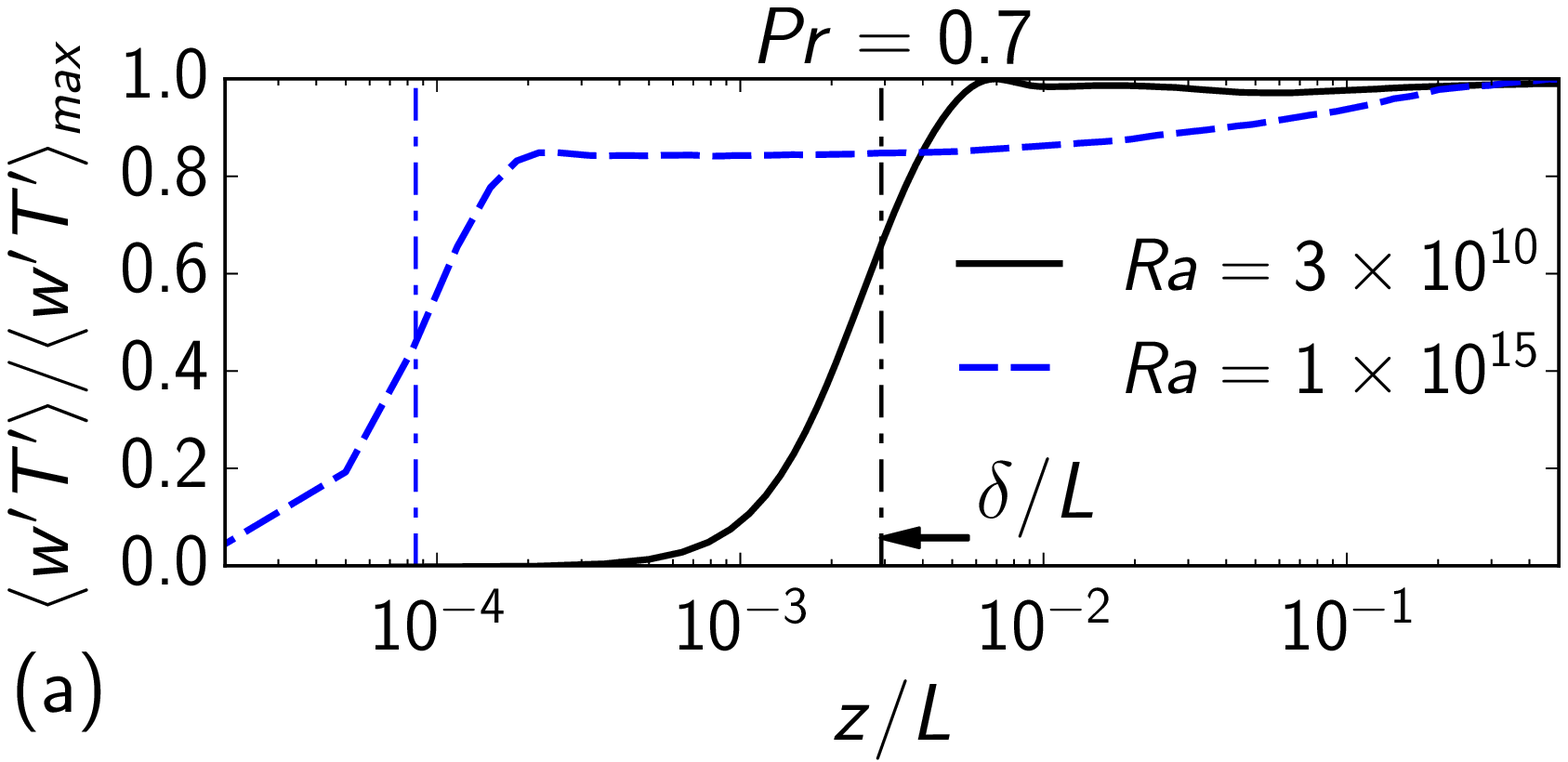}
  \includegraphics[width=67mm]{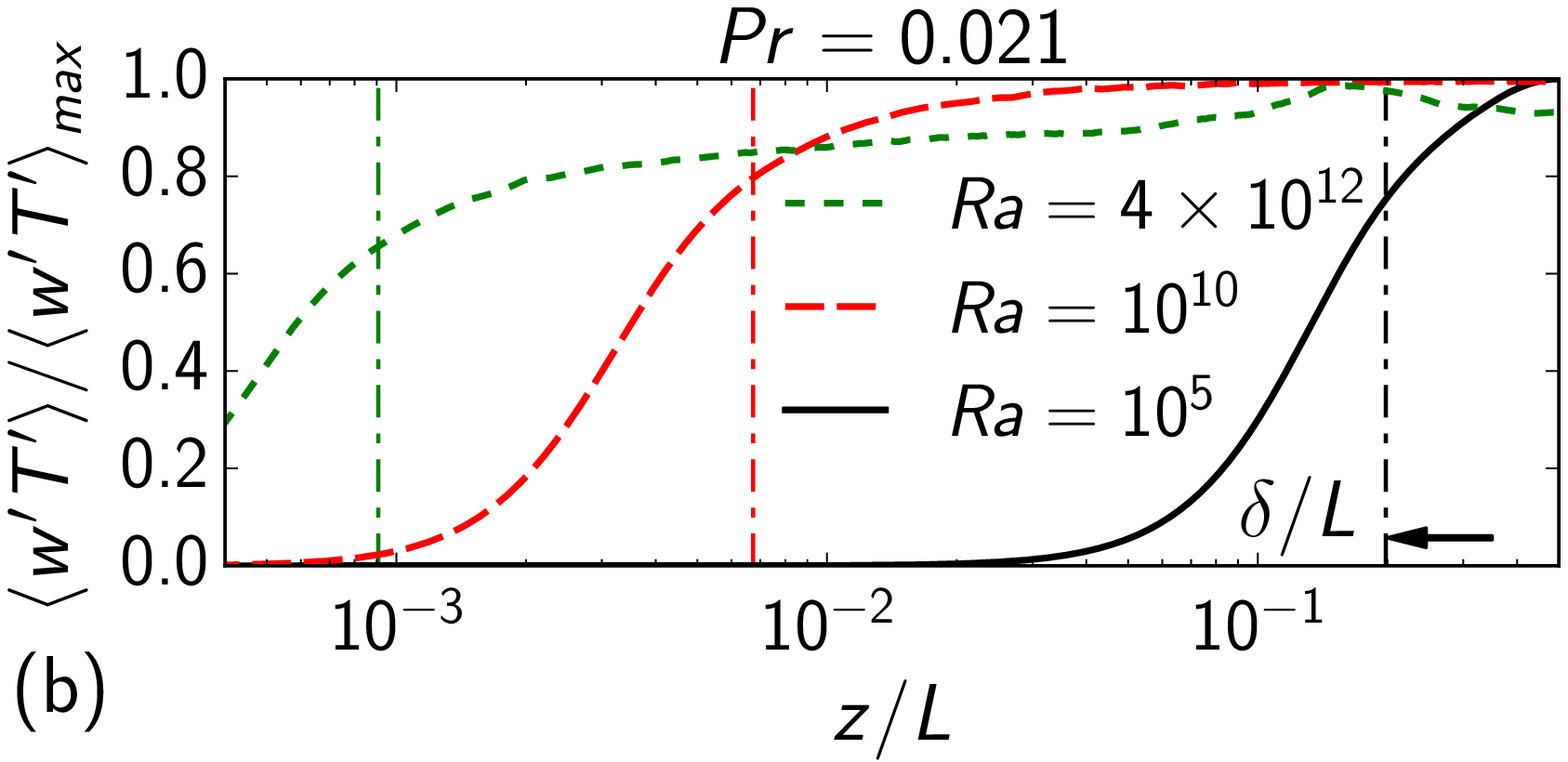}
  \caption{%
    Temperature-velocity cross-correlations $\langle w'T'\rangle$ in the lower half of the domain for (a) $Pr=0.7$ and (b)~$Pr=0.021$.
    Vertical lines mark the thermal boundary layer thickness $\delta$ as in figure~\ref{fig:BLprof1}.
  }
  \label{fig:BLflux}
\end{figure}

Figure~\ref{fig:BLflux} shows vertical profiles of the temperature-velocity cross-correlations $\langle w'T'\rangle$ obtained with ODT for the lower half of the domain.
The data have been normalised with the maximum value $\langle w'T'\rangle_{max}$ in order to focus on the shapes.
In the classical regime, for $Pr=0.7$, $Ra=3\times10^{10}$ in figure~\ref{fig:BLflux}(a), $\langle w'T'\rangle$ increases rapidly across the conductive sub-layer so that the maximum value is reached at $z/L\approx2\delta/L\approx6\times10^{-3}$.
This is similar for $Pr=0.021$, $Ra=10^{10}$ in figure~\ref{fig:BLflux}(b).
By contrast, the case $Pr=0.021$, $Ra=10^{5}$ in figure~\ref{fig:BLflux}(b), is dominated by thermal diffusion since the conductive sub-layer extends up to $z/L=2\delta/L\approx0.4$ and $\langle w'T'\rangle$ peaks at mid-height. 

In the ultimate regime, for $Pr=0.7$, $Ra=10^{15}$ in figure~\ref{fig:BLflux}(a), $\langle w'T'\rangle$ has only reached $\approx82\%$ of its maximum value at $z/L=2\delta/L\approx2\times10^{-4}$.
The temperature-velocity cross-correlation increases further with distance, but more gradually, and attains its maximum value in the bulk for $z/L\gtrsim0.18$.
A similar behaviour can be discerned for $Pr=0.021$, $Ra=4\times10^{12}$ in figure~\ref{fig:BLflux}(b), where $\langle w'T'\rangle$ reaches only $\approx75\%$ of its maximum value at $z/L=2\delta/L\approx2\times10^{-3}$. 
Likewise, the maximum temperature-velocity cross-correlation obtained with ODT is reached around $z/L\approx0.18$, but reduces again towards the mid-height.
This effect is robustly observed also for the higher $Ra$ numbers investigated. 

We note that the shape differences exhibited by $\langle w'T'\rangle$ for $Pr=0.7$, $Ra=3\times10^{10}$ and $Pr=0.021$, $Ra=10^{10}$ across the interval $10^{-3}\leq z/L\leq10^{-2}$ are related to the map-based advection representation analogously to the discussion of figure~\ref{fig:BLprof1} above.

\section{Conclusion} 
\label{sec:conc}

One-dimensional turbulence (ODT) simulations of high-$Ra$ thermal convection were conducted for $Pr=0.7$ and $0.021$, respectively, in order to numerically address the transition to the ultimate regime.
ODT is a stochastic turbulence model that aims to resolve all relevant scales of the turbulent flow for a wall-normal vertical line.
This model effectively mimics the direct cascade of featureless Kolmogorov turbulence within the dimensionally-reduced setting.
Here we have considered a Boussinesq fluid confined between smooth isothermal no-slip walls for a configuration with infinite aspect ratio.

For fixed $Pr$ and ODT parameters, the ODT model captures the classical $Nu\propto Ra^{1/3}$ scaling and the onset of the ultimate regime by realising $Nu\simeq \tilde{a}(Pr)\,Ra^{1/2}\left[\ln(Ra)\right]^{-3/2}$.
In the classical regime, ODT only slightly overestimates the scaling exponent of the available reference data.
The transition to the ultimate regime is observed for the critical Rayleigh numbers $Ra_*\simeq6\times10^{14}$ ($Pr=0.7$) and $Ra_*\simeq6\times10^{11}$ ($Pr=0.021$), respectively.
These values are within the ranges given in the literature but close to the upper limits \citep{He_etal:2012, Schumacher_etal:2016, Ahlers_etal:2017}.
In the ultimate regime, ODT results are remarkably well described by the \citet{Kraichnan:1962} theory.
This includes the prefactor $\tilde{a}$ for $Pr=0.7$, but not for $Pr=0.021$ for which it is an order of magnitude too small.
The reason for this discrepancy is unclear since otherwise the ODT results are consistent with the available reference data.
The slightly different classical scaling and the late transition are presumably related to unresolved large-scale motions.

At last, we note that the ODT results exhibit logarithmic temperature profiles prior and after the transition to the ultimate regime for large $Ra$ numbers.
This is in agreement with the literature \citep{Ahlers_etal:2012}.
After the transition, however, ODT yields a relative enhancement of the temperature-velocity cross-correlations in the bulk of the fluid.
This gives numerical support to an analogous assumption of \citet{Kraichnan:1962}.

\section*{Acknowledgements}

We thank Ambrish Pandey for providing DNS reference data.
Financial support by the European Regional Development Fund (EFRE), Grant No.\ {StaF}\ 23035000, and the German Academic Exchange Service (DAAD), which is funded by the Federal Ministry of Education and Research (BMBF), Grant No.\ {ID}-57316240, is kindly acknowledged.



\end{document}